\documentclass[superscriptaddress,amsmath,amssymb,twocolumn,pra,floatfix]{revtex4-1}
\usepackage{amsmath,amssymb}
\usepackage[usenames]{color}
\usepackage{mathbbol}
\usepackage{amssymb}
\usepackage{grffile}
\usepackage{amsmath, amstext, amssymb, amsfonts, amsxtra}
\usepackage{textcomp}
\usepackage{xspace}
\usepackage{hyperref}
\usepackage{enumitem}
\usepackage{bbold}

\newcommand{\ket}[1]{|\!\! #1 \rangle}
\newcommand{\bra}[1]{\langle #1 \!\!|}

\newcommand{\aver}[1]{\langle #1\rangle}

\newcommand{\im}{{\rm i}}

\newcommand{\figref}[1]{Fig.\ref{#1}}
\newcommand{\Ham}{\hat{\mathcal{H}}}
\newcommand{\sop}{\hat{\sigma}^+}
\newcommand{\som}{\hat{\sigma}^-}
\newcommand{\sz}{\hat{\sigma}^z}
\newcommand{\Dt}{\bar{\Delta}_L}

\begin{document}
\title{Giant rectification in segmented, strongly interacting, spin chains despite the presence of perturbations}

\author{Kang Hao Lee}
\affiliation{Science and Math Cluster, Singapore University of Technology and Design, 8 Somapah Road, 487372 Singapore}

\author{Vinitha Balachandran}
\affiliation{Science and Math Cluster, Singapore University of Technology and Design, 8 Somapah Road, 487372 Singapore}

\author{Dario Poletti}
\email{dario\_poletti@sutd.edu.sg}
\affiliation{Science and Math Cluster, Singapore University of Technology and Design, 8 Somapah Road, 487372 Singapore}
\affiliation{Engineering Product Development Pillar, Singapore University of Technology and Design, 8 Somapah Road, 487372 Singapore}

\date{\today}

\begin{abstract}
In Phys. Rev. Lett. 120, 200603 (2018), a segmented XXZ spin chain with zero anisotropy in one half and a large anisotropy on the other half gave rise to a spin current rectification which is perfect in the thermodynamic limit. Here we extend the previous study to segmented chains with interacting integrable as well as non-integrable halves, considering even cases in which no ballistic transport can emerge in either half. We demonstrate that, also in this more general case, it is possible to obtain giant rectification when the two interacting half chains are sufficiently different. We also show that the mechanism causing this effect is the emergence of an energy gap in the excitation spectrum of the out-of-equilibrium insulating steady state in one of the two biases. Finally we demonstrate that in the thermodynamic limit there is no perfect rectification when each of the two half chains is interacting.
\end{abstract}

\maketitle

\section{\label{sec:level1}Introduction}

The interplay between tunneling and interaction terms in spin chains can result in a large variety of transport properties \cite{BertiniZnidaric2020}. Taking simply an XXZ chain coupled to local magnetization baths, one can observe ballistic transport, super-diffusion, diffusion, sub-diffusion and an insulating behavior \cite{znidaric_spin_2011, prosen_open_2011}. The presence of symmetries and integrals of motions plays an important role in the emerging transport properties.
For instance, the presence of an extensive number of quasi-local conserved quantities with non-zero overlap with the current operator results in ballistic transport \cite{Mazur1969, Suzuki1971, prosen_open_2011, ZotosPrelovsek1997}. At the same time, non-integrability often results in diffusive transport, with notable exceptions \cite{Mastropietro2013, Znidaric2020, SchulzZnidaric2020, ZnidaricVarma2016, Znidaric2013, ZnidaricVipin2016,Brenes2018}.

Interestingly, for parameters close to the insulating transport regime, one can observe negative differential conductivity, i.e. the current can be reduced while increasing the bias difference between the two edges of the system \cite{benenti_negative_2009, benenti_charge_2009}.
Another interesting effect that can emerge in spin chains is that of current rectification, i.e. the system is such that one can observe very different current magnitudes when inverting the bias imposed by the baths \cite{ArracheaAligia2009, Landi2015, Schuab2016, balachandran_perfect_2018, BalachandranPoletti2019a,LeePoletti2019,Landi2014,Zhang2009,Werlang2014,Pereira2017,Mascarenhas2019}.

Rectification occurs in boundary driven quantum systems with broken reflection symmetry. However breaking of this symmetry is not sufficient to obtain rectification, and one would also need, for instance, the use of energy-dependent density of states for the baths, the use of different particle statistics \cite{WuSegal2009, WuSegal2009b}, the presence of nonlinearities in the system \cite{TerraneoCasati2002, LiCasati2004, LiLi2012} or temperature gradients within the baths \cite{Bijay2021}. Particularly strong rectification has been observed in systems with an excitation gap in reverse bias but not in forward bias \citep{balachandran_perfect_2018}, in the presence of mobility edges \cite{BalachandranPoletti2019}, negative differential conductance \cite{LeePoletti2020}, or phase transitions \cite{SchallerCelardo2016}.
In setups in which the baths, and their coupling to the system, are identical in nature, rectification is due to the properties of the system. In this scenario, the understanding of current rectification received a significant boost with the study of rectification of heat transport in classical nonlinear chains \cite{TerraneoCasati2002,LiCasati2004}. Strong rectification can emerge when the power spectra of two halves of the chain do not overlap for the bias in one direction, but they overlap for opposite bias. This physical principle can give strong rectification also for propagation of waves \cite{LepriCasati2011}.
From an experimental point of view, we point out that electrical currents rectification is pervasive in today's technology thanks to the works \cite{Starr1936, Brattain1951}, and important experiments on heat rectification have already been implemented \cite{Kouchmeshky2007a, Hu2009,ScheibnerMolenkamp2008, BoursGiazotto2019, FornieriGiazotto2015, IorioGiazotto2021,MartinezPerezGiazotto2015}.

This manuscript is particularly connected with a previous work by some of us, \cite{balachandran_perfect_2018}, where we considered a spin chain segmented in two parts, each with different anisotropy. This spin chain was coupled to two baths of different magnetization at the edges to study the spin current in the forward or reverse bias.
In such a system, transport can be diffusive when the baths are driving the current in one direction, and, in the thermodynamic limit, the system can become an insulator when swapping the baths, thus resulting in a perfect rectifier. For small system sizes, e.g. only $8$ spins, the current in one direction can be $10^4$ times larger than in the other.
In \cite{balachandran_perfect_2018} it was shown that the mechanism at the origin of current rectification is a strong mismatch in the excitation spectrum of spin excitations in one half compared to the other (analogous to the works on classical nonlinear chains \cite{TerraneoCasati2002,LiCasati2004}). However, in \cite{balachandran_perfect_2018} each half chain was integrable and, in particular, one half was non-interacting and hence ballistic. The occurrence of ballistic transport can be thought to have a critical role in the emergence of strong rectification. It is thus important to study whether strong, and even perfect, rectification can emerge in segmented spin chains under more general conditions for which each half may have interactions, and/or integrability breaking perturbations which can make both halves diffusive.

This is what we do in this work. First, we consider general segmented chains with interacting integrable halves. Then, we add two different integrability breaking perturbations on each half of the segmented XXZ chain. We show that, even in this most general scenario, it is still possible to obtain giant rectification when there is a large enough interaction strength difference between the two halves. To be more precise, we use the term giant rectification to indicate strong rectification emerging from a qualitatively different transport property in the forward or reverse bias. Specifically here we would encounter the case in which the system is diffusive in one bias, but insulating in the other. We show that the key to such behavior is still the emergence of a large excitation gap which can stabilize an insulating steady state in one of the two biases, while such excitation gap does not exist for the opposite bias.

The rest of the paper is structured as follows: in Sec.\ref{sec:model} we present the segmented model to be analyzed, in Sec.\ref{sec:results} we study the role of interactions followed by the role of integrability breaking perturbations in altering transport properties of the system. We show that giant rectification can be obtained in segmented chains provided there is a gap in the excitation spectra of two segments of the chain. We demonstrate this also using the Fourier transform of the current-current correlation function. The fact that one portion of the chain becomes almost an insulator is also evidenced by the magnetization profile. Importantly, we show that the magnitude of the interaction at which the gap opens is, when both portions of the chain have nonzero interaction, size dependent. The consequence is that in the thermodynamic limit there is no finite critical interaction strength beyond which the system becomes an insulator and hence a perfect rectifier. In Sec.\ref{sec:conclusions} we draw our conclusions.

\section{Model}\label{sec:model}
In the following we consider XXZ spin-1/2 chains with extra terms to break integrability. The total Hamiltonian is given by
\begin{align}\label{htotal}
\Ham &=\Ham_0+\Ham_{J_2}+\Ham_h.
\end{align}
where $\Ham_0$ is the base bipartite Hamiltonian in \cite{balachandran_perfect_2018}, while $\Ham_{J_2}$ and $\Ham_h$ are the two integrability breaking terms. For a chain with an even number $N$ of spins, $\Ham_0$ is given by
\begin{align}
\Ham_0 &= \sum_{n=1}^{N/2-1} [2J (\sop_n\som_{n+1} + \som_n\sop_{n+1}) + \Delta_L\; \sz_n\sz_{n+1}] \nonumber\\
&+ \sum^{N-1}_{n=N/2+1} [2J (\sop_n\som_{n+1} + \som_n\sop_{n+1}) + \Delta_R\; \sz_n\sz_{n+1}] \nonumber \\
&+2J_m (\sop_{N/2}\som_{N/2+1} + \som_{N/2}\sop_{N/2+1}).\label{baseHam}
\end{align}
Here $\sop_n$ and $\som_n$ are the raising and lowering operators acting on site $n$ and $\sz_n$ is a Pauli spin matrix, $J$ is the tunnelling amplitude, $\Delta_L$ and $\Delta_R$ are the anisotropy respectively in the left and right halves of the system, and $J_m$ is the tunnelling amplitude between the two half chains. Each half chain thus follows the so-called XXZ model, and in such model one can think of the anisotropy as interaction because of the correspondence, via Jordan-Wigner transformation \cite{JordanWigner1928, LiebMattis1961}, between this spin-chain and the spinless Fermi-Hubbard model with nearest neighbor interaction. In the following, we will thus freely interchange the terms anisotropy and interactions.
The model described by the Hamiltonian $\Ham_0$ is, overall, non-integrable, but each half of the chain is integrable and can have, depending on the magnitude of $\Delta_L$ or $\Delta_R$, ballistic spin transport (for $\Delta_{L},\Delta_{R}<J$) or diffusive transport (for $\Delta_{L},\Delta_{R}>J$).
This model was analyzed in \cite{balachandran_perfect_2018}, and it was shown that for $\Delta_R=0$ and $\Delta_L$ large enough, e.g. $\Delta_L/J\ge 1+\sqrt{2}$, one can obtain, in the thermodynamic limit, diffusive transport with one bias, and an insulating steady state with the opposite bias. Hence, one can realize a perfect rectifier.

The first integrability breaking term $\Ham_{J_2}$ is a  next-nearest neighbor tunneling and is described by ,
\begin{align}
\Ham_{J_2}&=\sum_{\substack{{n=1}\\ n \neq L/2-1,L/2}}^{N-2} [2J_2(\hat{\sigma}^+_n\hat{\sigma}^-_{n+2} + \hat{\sigma}^-_n\hat{\sigma}^+_{n+2}),
\label{nnn}
\end{align}
where $J_2$ is the amplitude of the next nearest neighbor tunneling. The next integrability breaking term $\Ham_h$ is
 a staggered magnetic field with amplitude $h$ and is given by
\begin{equation}
\begin{split}
\Ham_{h}&= h\sum_{n=1}^{N} (-1)^{n-1}\hat{\sigma}^z_n.
\end{split}
\label{stagfield}
\end{equation}

The chain is coupled to two spin baths at the edges and we model the evolution via a Gorini - Kossakowski - Sudarshan - Lindblad (GKSL) master equation \cite{gorini_completely_1976,lindblad_generators_1976}
\begin{equation}
\frac{d\hat{\rho}}{d t}=-\frac{\rm{i}}{\hbar}[\hat{\mathcal{H}},\hat{\rho}]+\sum^4_{n=1}D_n\hat{\rho} D_n^\dagger-\frac{1}{2}\sum^4_{n=1}\{D^\dagger_nD_n,\hat{\rho}\},
\label{lindblad}
\end{equation}
where the $D_n$ are the jump operators given by
\begin{equation}
\begin{aligned}
    D_1 &= \sqrt{\Gamma\mu_L}\hat{\sigma}_1^+, & D_2 &=\sqrt{\Gamma(1-\mu_L)}\hat{\sigma}_1^-,\\
    D_3 &= \sqrt{\Gamma\mu_R}\hat{\sigma}_N^+, & D_4 &=\sqrt{\Gamma(1-\mu_R)}\hat{\sigma}_N^-.
\end{aligned}
\label{eq: set of dissipators}
\end{equation}
Here $\Gamma$ is the magnitude of the system-reservoir coupling while the spin magnetization imposed by the left (right) bath is set by $\mu_L$ ($\mu_R$). We focus on $\mu_L$ and $\mu_R$ to be equal to 0 or 0.5. The imbalance $\mu_L-\mu_R$ induces a spin current. We refer to the spin current flowing from left to right ($\mu_L=0.5>\mu_R=0$) as forward bias, where the left bath tries to impose the spins to be in an equal mixture of up and down spins $(\ket{\downarrow}_n\bra{\downarrow}+\ket{\uparrow}_n\bra{\uparrow})/2$ while the right bath tends to set the spins pointing down $\ket{\downarrow}_n\bra{\downarrow}$. The reverse bias thus corresponds to the spin current flowing from right to left ($\mu_L=0<\mu_R=0.5$). Effects of using different values from $0$ and $0.5$ for $\mu_L$ and $\mu_R$ where analyzed already in \cite{balachandran_perfect_2018}, showing a reduction of the rectification.

The spin current $\mathcal{J}$ is calculated from the continuity equation for local observable $ \hat{\sigma}_n^z$ and is given by
\begin{equation}
\mathcal{J} = \textrm{Tr}(\hat{j}_n \hat{\rho}_{ss}), \label{eq:steadycurrent}
\end{equation}
where $\hat{\rho}_{ss}$ is the steady state such that $d\hat{\rho}_{ss}/dt=0$ and
\begin{equation}
\begin{aligned}
\hat{j}_{n}  &= \frac{4\rm{i}}{\hbar}[J(\hat{\sigma}^+_n \hat{\sigma}^-_{n+1} - \hat{\sigma}^-_n \hat{\sigma}^+_{n+1})\\
&+ J_2(\hat{\sigma}^+_n \hat{\sigma}^-_{n+2} - \hat{\sigma}^-_n \hat{\sigma}^+_{n+2})].
\end{aligned}
\label{current}
\end{equation}
Note that for $n = N/2$ and $N/2-1$, the second term in Eq. (\ref{current}) is not considered because those sites are not connected by the next-nearest neighbor coupling $J_2$. Also, we note that the spin current $\mathcal{J}$ is site independent in the steady state, and for this reason $\mathcal{J}$ in Eq.(\ref{eq:steadycurrent}) has no index $n$.

To evaluate the performance of the system as a diode, we consider the measure $\mathcal{R}$ \cite{balachandran_perfect_2018}
\begin{equation}
\mathcal{R}=-\frac{\mathcal{J}_f}{\mathcal{J}_r},
\label{eq: contrast}
\end{equation}
where $\mathcal{J}_f$ and $\mathcal{J}_r$ are the currents in forward and reverse bias respectively.
When there is no rectification, $\mathcal{J}_f=-\mathcal{J}_r$ and the measure $\mathcal{R}=1$. For a perfect diode, $\mathcal{R}= \infty$ or $0$ because either the forward or reverse current goes to zero.

\section{Results}\label{sec:results}
\subsection{Case of homogeneous interaction within each segment}\label{sec:ixn}
We study the spin current rectification of a general segmented chain using exact diagonalization with a significant reduction of the memory requirements by implementing a method that takes into account of the fact that the Hamiltonian is magnetization-conserving, while the baths can only change the spins at the edges in the form described by Eqs.(\ref{lindblad},\ref{eq: set of dissipators}). This method was first proposed and implemented in \cite{guo_dissipatively_2017} where it allowed to compute the quantum open system dynamics up to $14$ spins.

\begin{figure}[ht]
    \includegraphics[trim=0 0 0 0,clip, width=\linewidth]{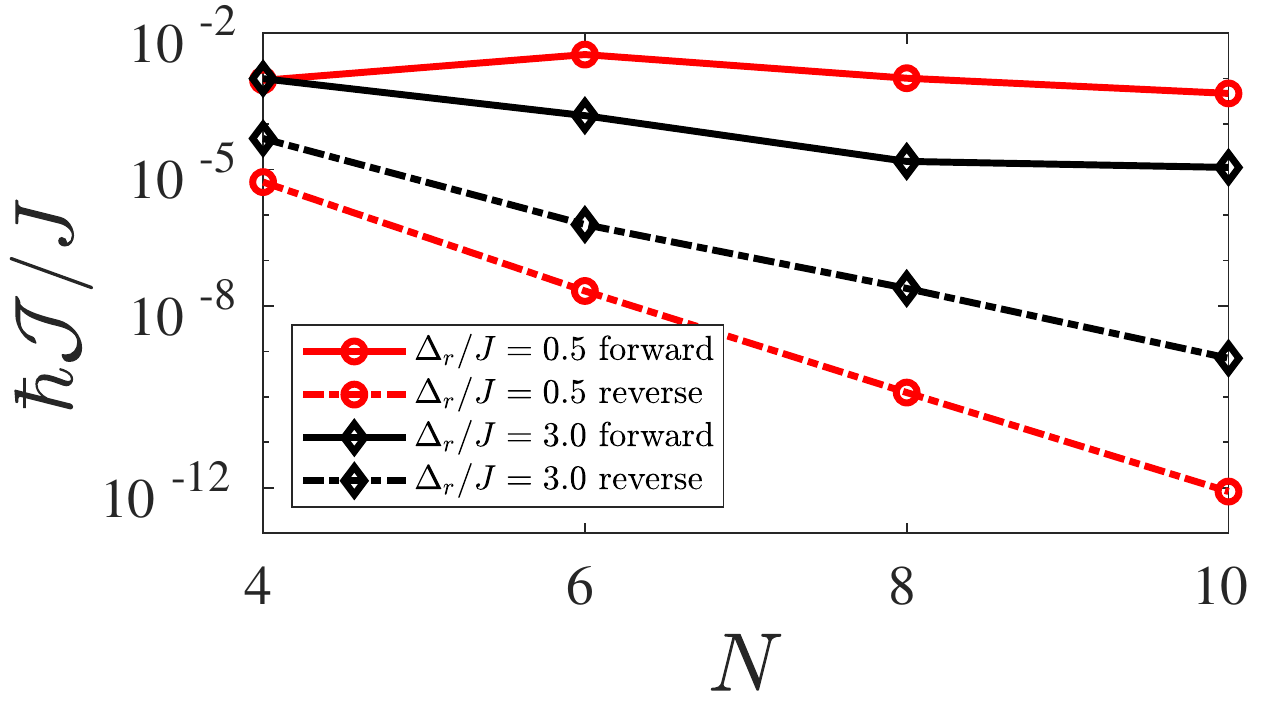}
\caption{Forward (continuous lines) and reverse (dot-dashed lines) currents $\mathcal{J}$ vs system size $N$ for a fixed left anisotropy $\Delta_L/J=15$ and right anisotropies $\Delta_R/J=0.5$ (empty red circles) and $\Delta_R/J=3.0$ (empty black diamonds). Other parameters: $J_2=0$, $h=0$, $J_m=0.1J$, $\Gamma=J/\hbar$. }
\label{fig:scaling}
\end{figure}

For giant rectification to occur in this bipartite system, the transport properties when the bias is applied in opposing directions should be qualitatively different. This was observed, for instance, in \cite{balachandran_perfect_2018} where the transport was diffusive in forward bias  and insulating in the reverse bias. However, in \cite{balachandran_perfect_2018} one half of the chain was non interacting, and it is thus natural to investigate whether such phenomenology occurs also when both halves are interacting.
We start our investigation discussing the dependence of the current on the system size. For an insulator the current should decay exponentially, while in diffusive, subdiffusive or superdiffusive scenarios, it would follow a power-law.
In Fig.\ref{fig:scaling} we consider the current decay in forward (continuous lines) or reverse bias (dot-dashed lines) for $\Delta_R=0.5J$ (red lines with circles) and for $\Delta_R=3.0J$ (black lines with diamonds). The choice of these two different values of $\Delta_R$ is motivated by the fact that the right half of the chain would have ballistic transport for $\Delta_R<J$ and diffusive for $\Delta_R>J$.
In both cases, $\Delta_L=15J$ which, as we will show later, is sufficient to have giant rectification for both the values of $\Delta_R$ considered.
We observe that in both cases the dependence of the current with the system size is an exponential decay in reverse bias, and it is a significantly slower decay (hinting at an algebraic decay), in forward bias. This is a first indication of the emergence of qualitatively different transport properties in the two biases, which leads to giant rectification. Importantly, we do not observe qualitative differences depending on whether $\Delta_R$ is greater or lesser than $J$.
We note that there is a strong limitation to the system sizes we can consider, and this is due to the fact that for the model and parameters considered, there exist decay modes with an extremely long relaxation time scale which are difficult to differentiate from the unique steady state.
\begin{figure}[ht]
    \includegraphics[trim=0 0 0 0,clip, width=\linewidth]{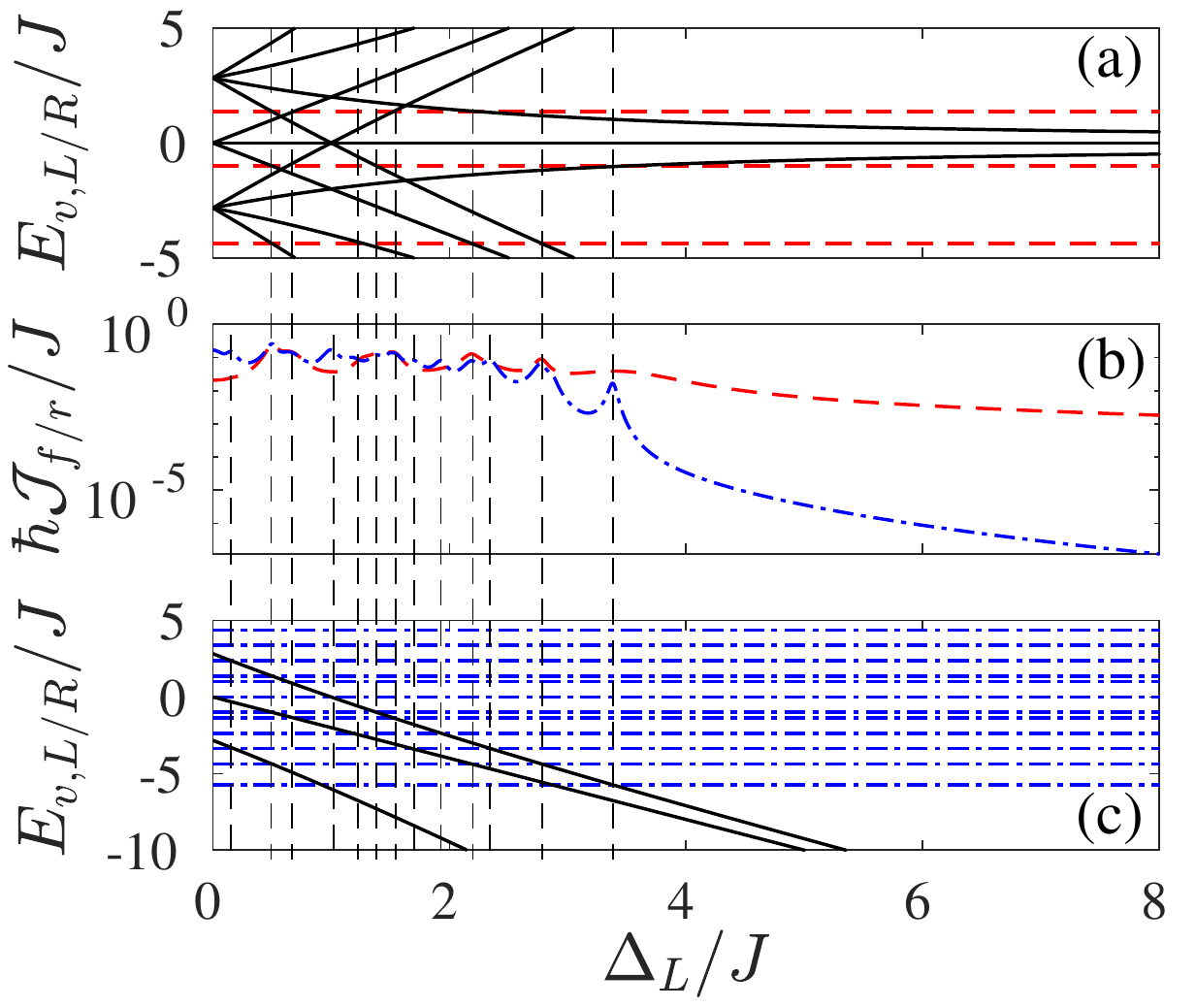}
\caption{Excitation spectra $E_{v,L/R}$  and current $\mathcal{J}$ for a fixed right anisotropy $\Delta_R/J=0.5$ against left anisotropy $\Delta_L$. (a) For the forward bias scenario, excitation spectra $E_{v,L}$ (black continuous lines) and $E_{v,R}$ (red dashed lines) as a function of anisotropy $\Delta_L/J$.  (b) Forward current $\mathcal{J}_f$ (red dashed lines) and reverse current $\mathcal{J}_r$ (blue dot-dashed lines) against left anisotropy $\Delta_L/J$. (c) For the reverse bias scenario, excitation spectra $E_{v,L}$ (black continuous lines) and $E_{v,R}$ (blue dot-dashed lines) as a function of anisotropy $\Delta_L/J$. Vertical dashed lines indicate peaks in the currents of panel (b) corresponding to  $E_{v,L} = E_{v,R}$ in panel (c) for reverse current and/or panel (a) for forward current. Other parameters: $N=6$, $J_2=0$, $h=0$, $J_m=0.1J$, $\Gamma=0.1J/\hbar$.} %\dpc{Is $N=6$ or $12$?} }
\label{fig:fig1_05enermatch}
\end{figure}

\begin{figure}[ht]
    \includegraphics[trim=0 0 0 0,clip, width=\linewidth]{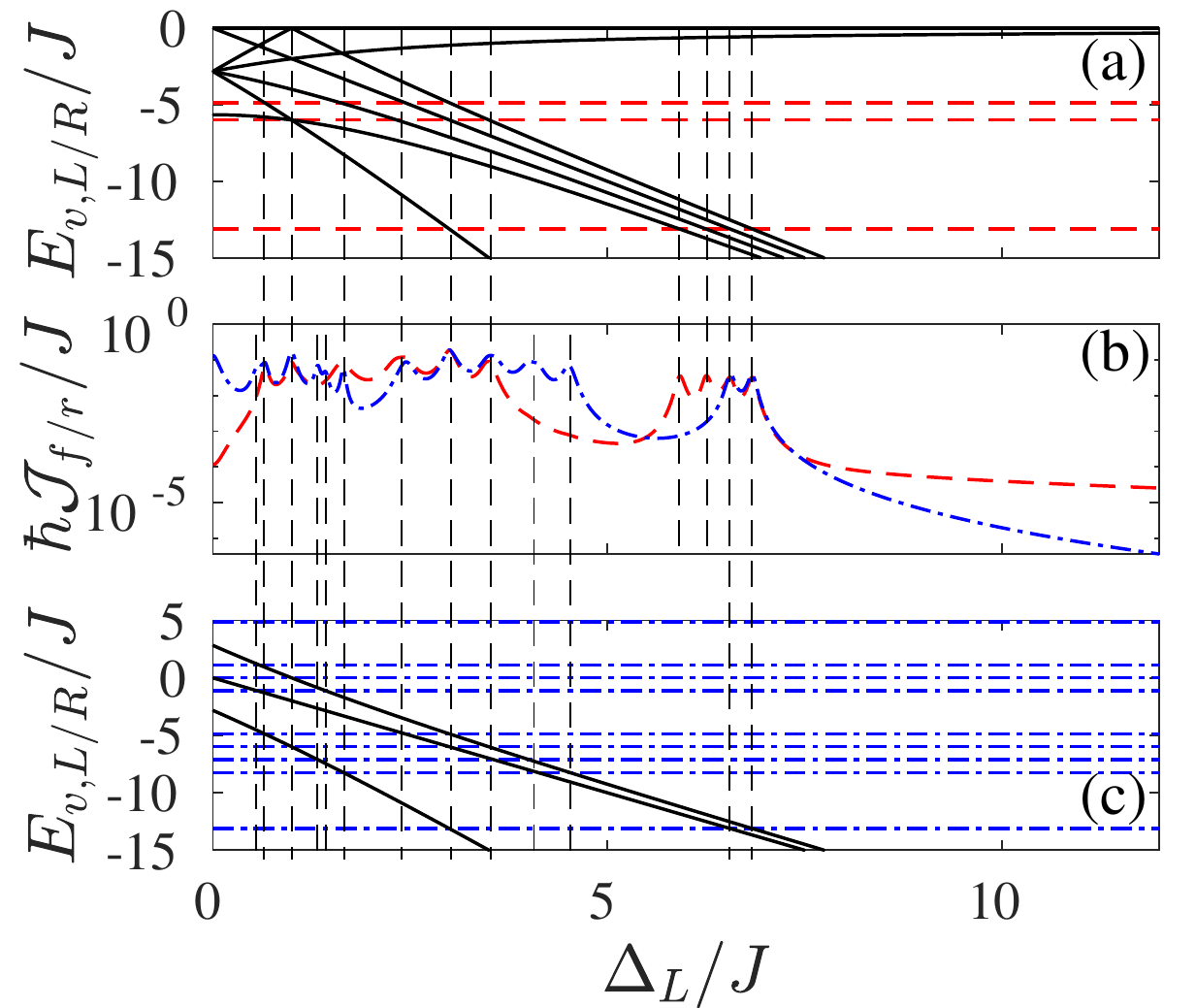}
\caption{Excitation spectra $E_{v,L/R}$  and current $\mathcal{J}$ for a fixed right anisotropy $\Delta_R/J=3.0$ against left anisotropy $\Delta_L$. (a) For the forward bias scenario, excitation spectra $E_{v,L}$ (black continuous lines) and $E_{v,R}$ (red dashed lines) as a function of anisotropy $\Delta_L/J$.  (b) Forward current $\mathcal{J}_f$ (red dashed lines) and reverse current $\mathcal{J}_r$ (blue dot-dashed lines) against left anisotropy $\Delta_L/J$. (c) For the reverse bias scenario, excitation spectra $E_{v,L}$ (black continuous lines) and $E_{v,R}$ (blue dot-dashed lines) as a function of anisotropy $\Delta_L/J$. Vertical dashed lines indicate peaks in the currents of panel (b) corresponding to  $E_{v,L} = E_{v,R}$ in panel (c) for reverse current and/or panel (a) for forward current. Other parameters: $N=6$, $J_2=0$, $h=0$, $J_m=0.1J$, $\Gamma=0.1J/\hbar$. }%\dpc{Is $N=6$ or $12$?} }
\label{fig:fig2_3enermatch}
\end{figure}
We then consider the dependence of the current versus anisotropy in the left half $\Delta_L$ in \figref{fig:fig1_05enermatch}(b), for $\Delta_R=0.5J$, and \figref{fig:fig2_3enermatch}(b), for $\Delta_R=3J$.
In both figures we observe that beyond a certain value of $\Delta_L$, which is different in each figure, the reverse bias current (blue dot-dashed line) decays much faster than the forward bias current (red dashed line). In the following we will refer to this threshold value of anisotropy as $\Dt$. It is for these large values of $\Delta_L$ that the reverse bias current decay exponentially with the system size, or in other words, the system becomes an insulator.
We also highlight to the reader that the $y-$axis, in both \figref{fig:fig1_05enermatch}(b) and \figref{fig:fig2_3enermatch}(b), is in log scale, which implies that the difference between the currents in forward and reverse bias is of various orders of magnitude, and hence the rectification is extremely large.

For $\Delta_L<\Dt$, we observe several peaks in the forward as well as in the reverse currents. Similarly to the analysis done in \cite{balachandran_perfect_2018}, the formation of the peaks in both the currents can be understood by considering two chains, one fully polarized such that $\hat{\rho}_{\downarrow}=\otimes_n\ket{\downarrow}_n\bra{\downarrow}$ and the other chain at an infinite temperature state such that $\hat{\rho}_{\infty}=\otimes_n(\ket{\uparrow}_n\bra{\uparrow}+\ket{\downarrow}_n\bra{\downarrow})/2$. Note that these are the two states that the baths are trying to impose in each half of the chain if these two halves are not coupled. When the two halves are coupled together, spin excitations will form at the interface between the two halves. Such excitations will be localized at the interface if there is an energy gap between the excitation spectra of two halves, resulting in an insulating system. If there is no gap, excitations that form at the interface can propagate until they reach the baths and can be exchanged with them, thus resulting in a nonzero steady state current.
To determine the existence of an energy gap in the excitation spectrum we operate differently in the two halves because the initial condition is different: one is fully polarized, $\hat{\rho}_{\downarrow}$, while the other is in the infinite temperature state, $\hat{\rho}_{\infty}$. For the fully polarized half chain, the only possible excitations are single spin excitations. For the half chain in the infinite temperature state, we consider all the possible energy transitions, from one magnetization sector to a sector with one less spin up. For the forward bias case in panels (a) of \figref{fig:fig1_05enermatch} and \figref{fig:fig2_3enermatch}, the left half chain is in the infinite temperature state and the resulting excitations $E_{v,L}$ are depicted with continuous black lines whereas the right half chain is in the fully polarized state and the excitations $E_{v,R}$ are plotted in red dashed lines. For the reverse bias case in panels (c) of \figref{fig:fig1_05enermatch} and \figref{fig:fig2_3enermatch}, the left half chain is in the fully polarized state with excitations $E_{v,L}$ plotted in continuous black line and the right half chain in the infinite temperature state with excitations $E_{v,R}$ plotted in blue dot-dashed lines.

In reverse bias, panels \figref{fig:fig1_05enermatch}(c) and \figref{fig:fig2_3enermatch}(c), we observe clearly that beyond a certain $\Dt$, there is no overlap between the spectra, and the current will be more and more suppressed.
In forward bias, panels \figref{fig:fig1_05enermatch}(a) and \figref{fig:fig2_3enermatch}(a), there are excitations in the left half which are close to zero. Furthermore, for larger system sizes these excitations form a band and an overlap of the excitation spectra of the two half chains is possible.
We highlight that each peak in the currents, whether in forward or reverse bias, coincides with a match in the excitation energies in the left and right halves, because that is when zero energy excitation can propagate readily along the spin chain. We also stress that for larger systems, the excitation energy bands will be more densely filled, thus resulting in a continuum of large current for a (finite) range of values of $\Delta_L$.
This qualitative phenomenology occurs for both $\Delta_R$ larger or smaller than $J$, indicating that this is independent on the type of transport present in the right half of the chain.

\begin{figure}[ht]
    \includegraphics[trim=0 0 0 0,clip, width=\linewidth]{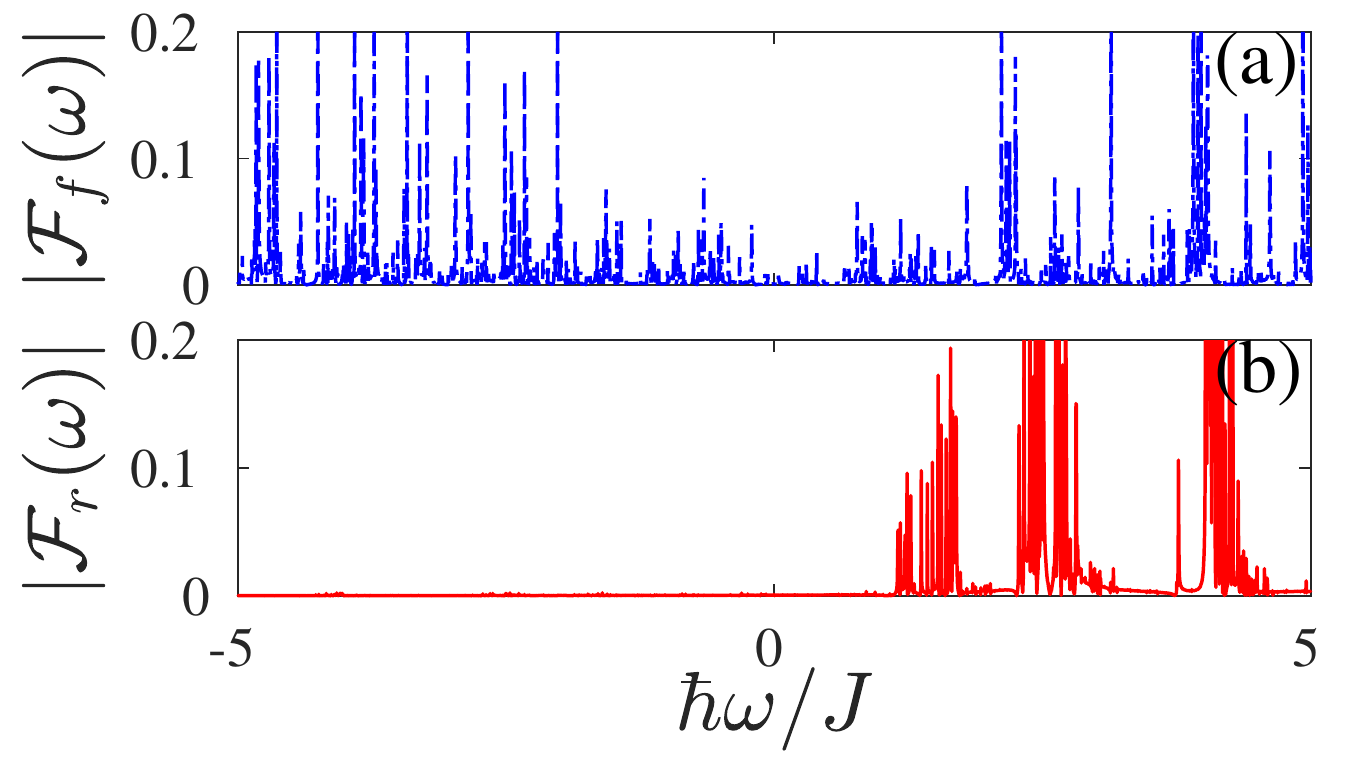}
\caption{Frequency response $|\mathcal{F}_{f/r}(\omega)|$ as a function of frequency $\hbar\omega/J$ for forward [blue dot-dashed lines in panel (a)] and reverse bias [red continuous lines in panel (b)] for a system size of $N=12$ with $\Delta_L/J=3$, $\Delta_R/J=0.1$, $J_2=0$, $h=0$ and $J_m=0.1J$. The computations were done up to a time $t'=2000 \hbar/J$ with $dt=0.05\hbar/J$.}
\label{fig:fig3_freqresp}
\end{figure}

We can gain deeper insight into the role of spin excitations by studying the dynamical correlations of a spin flip at the interface. More precisely, we study the frequency response of the tunneling excitation at the interface via the two-time correlation function
\begin{equation}
\mathcal{F}(t)=\mbox{tr}[ \hat{j}_{N/2}  e^{-\im\Ham t/\hbar}(\hat{j}_{N/2}\hat{\rho}_{\rm seg})e^{\im\Ham t/\hbar}]\hbar^2/J^2.
\label{eq: 2timecorr}
\end{equation}
where $\hat{\rho}_{\rm seg}$ is the steady state density matrix if $J_m=0$ for $\Ham$ defined in Eq.(\ref{htotal}). The Fourier transform of this two-time correlation is the frequency response $\mathcal{F}(\omega)$. In \figref{fig:fig3_freqresp}, we plot $\mathcal{F}(\omega)$ against $\omega$ for a system  of size $N=12$. We consider all spins down on one half of the chain coupled to the bath with $\mu=0$, $\hat{\rho}_{\downarrow}$, and for the other half coupled to the bath with $\mu=0.5$, we consider an equal mixture of up and down spins $\hat{\rho}_{\infty}$.

In the reverse bias, we thus take as initial condition $\hat{\rho}_{{\rm seg}, r}=\hat{\rho}_{\downarrow}\otimes\hat{\rho}_{\infty}$ whereas for the forward bias we choose $\hat{\rho}_{{\rm seg}, f}=\hat{\rho}_{\infty}\otimes\hat{\rho}_{\downarrow}$.
In \figref{fig:fig3_freqresp} we see clearly the emergence of an excitation gap in the reverse bias which is not present in the forward bias. Note that in order to compute the Fourier transform we have evolved the system with a time interval $dt=0.05 \hbar/J$ (which gives a bound to the largest frequencies we can observe), and over a time $t'=2000\hbar/J$ (which gives the resolution for small frequencies).

\begin{figure}[ht]
    \includegraphics[trim=0 0 0 0,clip, width=\linewidth]{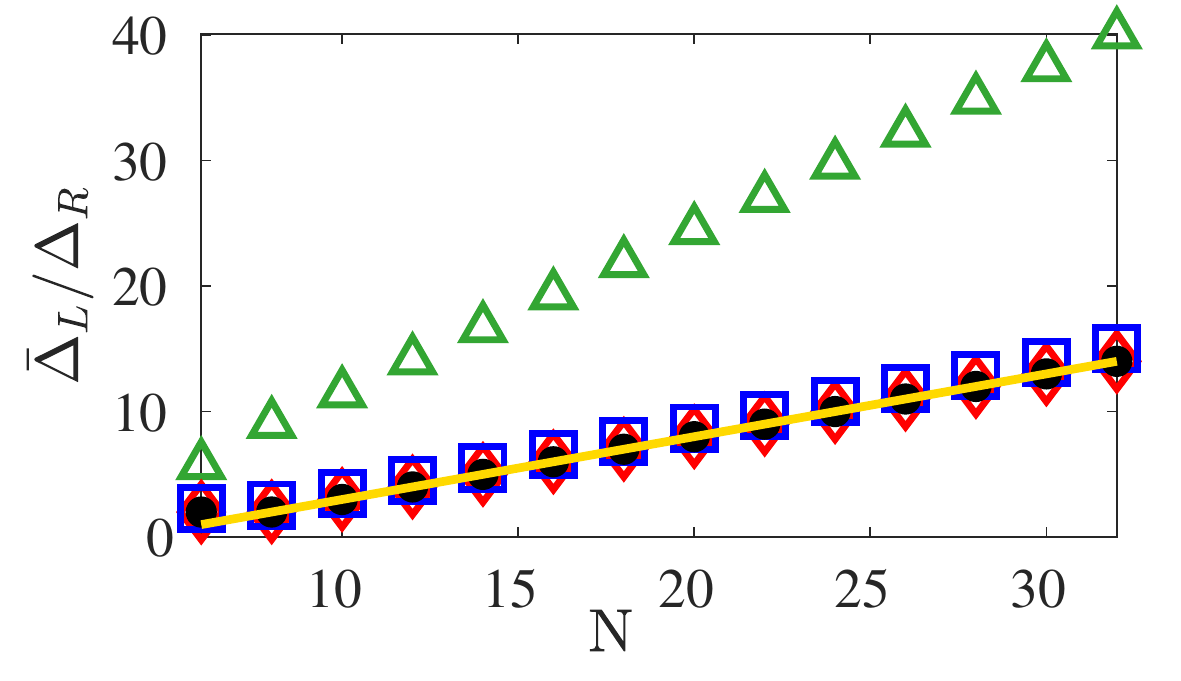}
\caption{Threshold anisotropy $\Dt/\Delta_R$ against system size $N$ for  $\Delta_R/J=0.5$ (empty green triangles) and $\Delta_R/J=3$ (empty blue squares). To showcase the limit cases of small $J$, we also consider $\Delta_R/J=50$ (empty red diamonds) and $\Delta_R/J=300$ (filled black circles). The yellow continuous line, which match the symbols (empty red diamonds and filled black circles), correspond to Eq.(\ref{eq:limit_case_threshold}).}
\label{fig:fig3_critpt}
\end{figure}

In \figref{fig:fig1_05enermatch} and \figref{fig:fig2_3enermatch} we have observed that, independent of whether $\Delta_R$ is smaller or larger than $J$, there is a threshold anisotropy beyond which an energy gap opens in the excitation spectrum and the system becomes an insulator. An important question now is whether there exist a critical value of anisotropy also in the thermodynamic limit or, more simply, whether the threshold anisotropy for the emergence of the insulating behavior is size dependent and tends to infinity for infinite-size systems. In Fig.\ref{fig:fig3_critpt} we plot the threshold anisotropy vs system size for $\Delta_R=0.5J$ (empty green triangles) and $\Delta_R=3J$ (empty blue squares).
We observe that the threshold anisotropy $\Dt$ increases with the system size. This is due to the fact that the width of the energy band in the right-half of the chain also increases with the system size.
To have a better insight into this, we consider the limit of $\Delta_{L},\Delta_{R} \gg J$, for which we obtain analytically the threshold anisotropy as
\begin{align}
\Dt=\Delta_R(N/2 - 2) \label{eq:limit_case_threshold}
\end{align}
The analytical form of $\Dt$, in the limit of strong interactions, is derived from analysing the energy changes due to a spin flip down from the right chain and a spin flip up on the left chain for the reverse bias case. For the right half of the chain coupled to the reservoir with $\mu_R=0.5$, the greatest difference in energy due to a spin flip lies between the antiferromagnetic state $\ket{\downarrow\uparrow\downarrow...\downarrow\uparrow\downarrow}$, and the state with two ferromagnetic domains that are oppositely polarised $\ket{\downarrow\dots\downarrow\uparrow\dots\uparrow}$, which is approximately $(N-4)\Delta_R$. Whereas in the left half of the chain coupled to the reservoir with $\mu_R=0$, the energy change will be due to a spin flip from the state $\ket{\downarrow\downarrow...\downarrow\downarrow}$ to $\ket{\downarrow...\downarrow\uparrow}$, which can be approximated by $2\Delta_L$.
Eq.(\ref{eq:limit_case_threshold}) is depicted with yellow continuous line in Fig.\ref{fig:fig3_critpt} and it accurately fits with numerical computations for $\Delta_R/J=50$ (empty red diamonds) or $\Delta_R/J=300$ (filled black circles). It is straightforward to use Eq.(\ref{eq:limit_case_threshold}) to show that there is no finite threshold $\Dt$ for an infinite-size system.
It is thus important to mark the main difference between the results in \cite{balachandran_perfect_2018}, where a perfect diode was observed in the thermodynamic limit, and the current work. The main point is that the energy bandwidth of a non-interacting chain is independent of the system size and guarantees a perfect diode, while in presence of interactions this does not have to occur.
We stress that, despite the lack of a perfect diode in the thermodynamic limit, giant rectification can still emerge, as indicated by the orders in magnitude difference between the currents in forward and reverse bias shown in panel (b) of Fig.\ref{fig:fig1_05enermatch} and \ref{fig:fig2_3enermatch}.

\subsection{Case of integrability-breaking perturbations}
\begin{figure}[ht]
    \includegraphics[trim=0 0 0 0,clip, width=\linewidth]{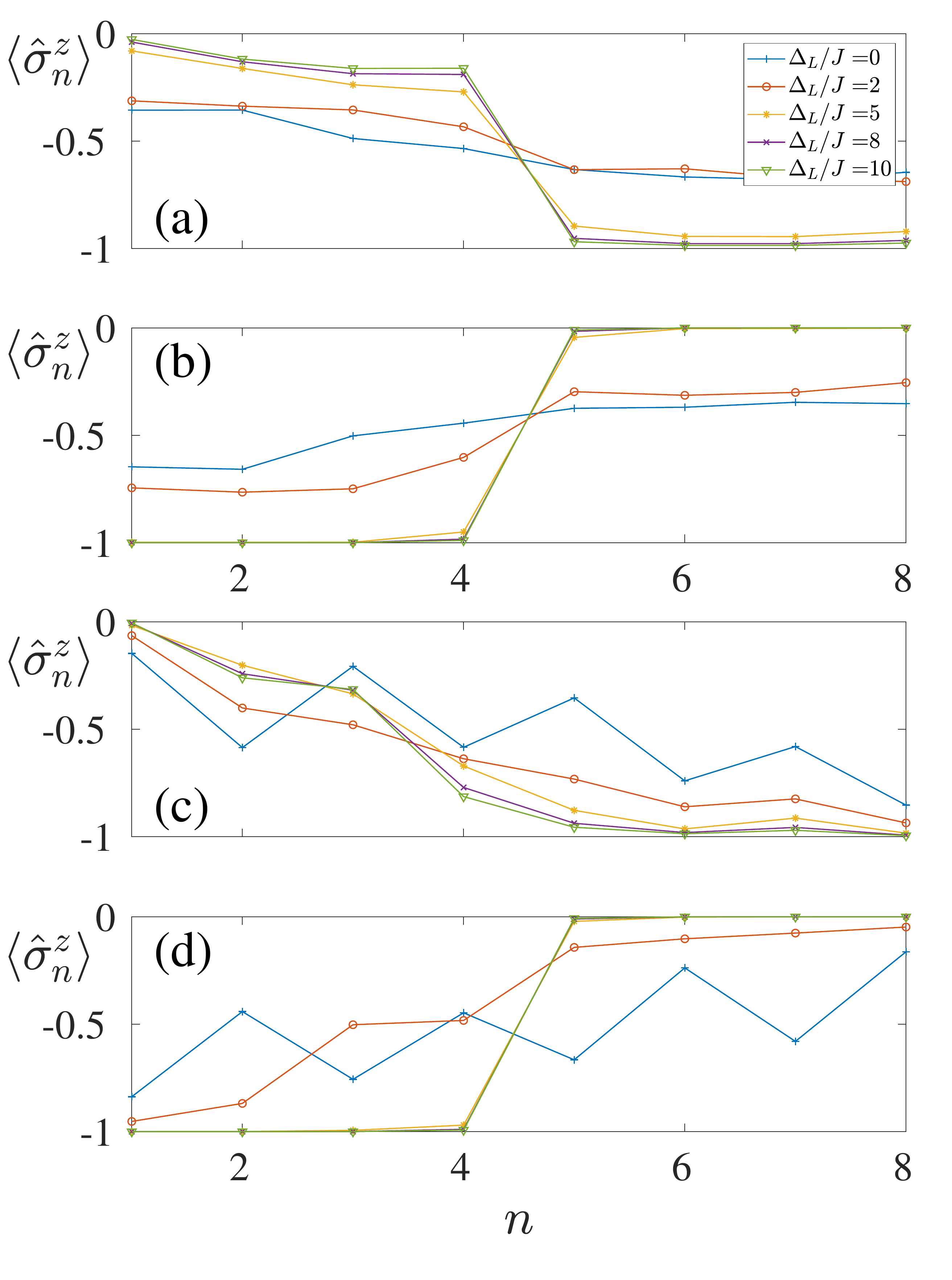}
\caption{(a, b) Profile of magnetization $\langle\hat{\sigma}^z_n\rangle$ as a function of the position index $n$ for a chain of $N=8$ spins with next-nearest-neighbour tunneling $J_2=0.5J$ and right anisotropy $\Delta_R/J=0.5$ for different values of left anisotropy $\Delta_L/J$. (a) Forward bias ($\mu_L=0.5$ and $\mu_R=0$) and (b) reverse bias ($\mu_L=0$ and $\mu_R=0.5$). Other parameters: $J_m=J$, $h=0$, $\Gamma=J/\hbar$. (c, d) Profile of magnetization $\langle\hat{\sigma}^z_n\rangle$ as a function of the position index $n$ for a chain of $N=8$ spins with staggered magnetic field perturbations $h=J$ and right anisotropy $\Delta_R/J=0.5$ for different values of left anisotropy $\Delta_L/J$. (c) Forward bias ($\mu_L=0.5$, $\mu_R=0$) and (d) reverse bias ($\mu_L=0$, $\mu_R=0.5$). Other parameters: $J_m=J$, $J_2=0$,  $\Gamma=J/\hbar$.}
\label{fig:sznnnstag}
\end{figure}

\begin{figure}
    \includegraphics[trim=0 0 0 0,clip, width=\linewidth]{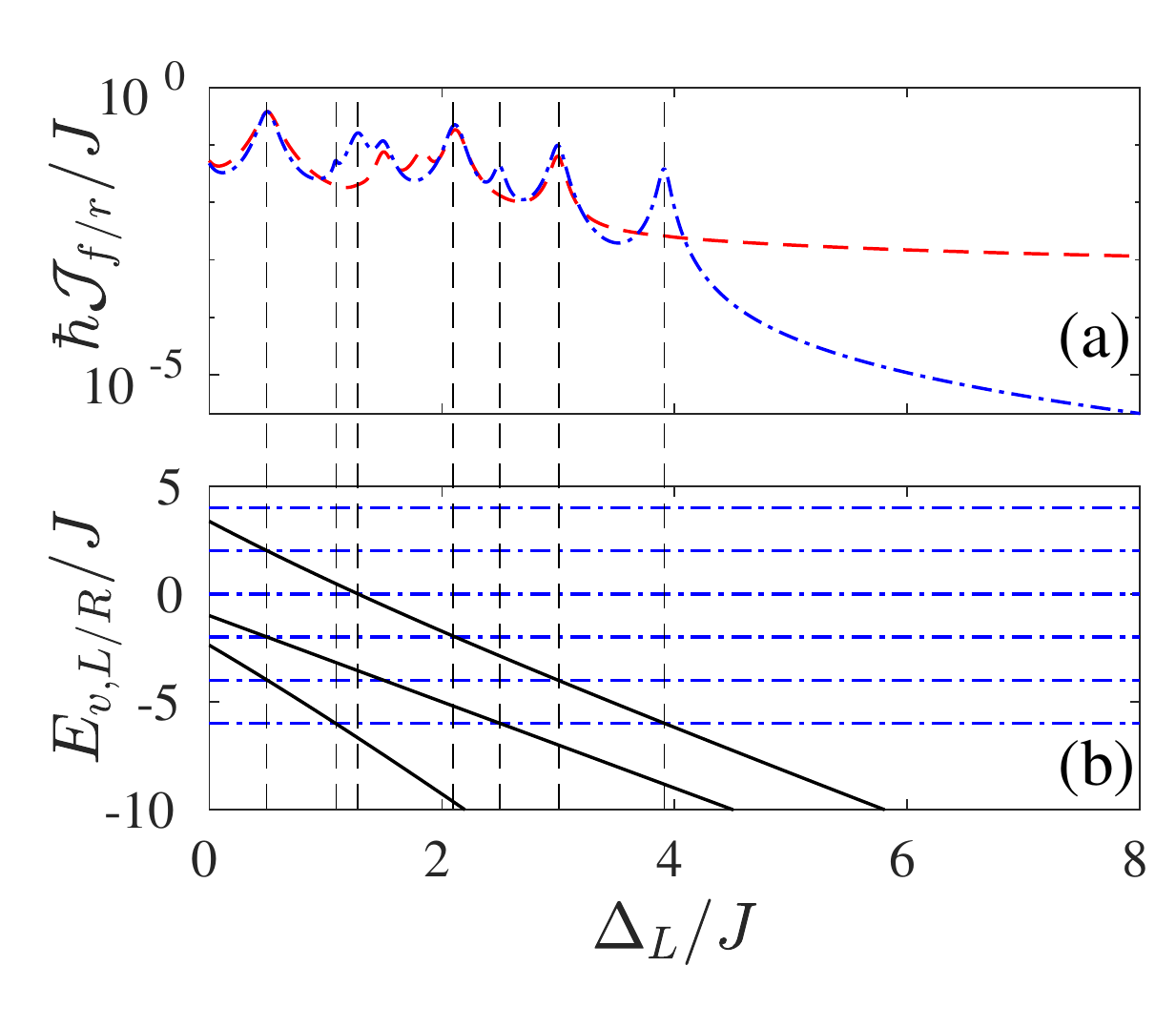}
\caption{Current $\mathcal{J}$ and excitation spectra $E_{v,L/R}$ against left anisotropy $\Delta_L$ with fixed next-nearest neighbour tunneling $J_2=0.5J$ and right anisotropy $\Delta_R/J=0.5$. (a) Forward current $\mathcal{J}_f$ (red dashed lines) and reverse current $\mathcal{J}_r$ (blue dot-dashed lines) against anisotropy on the left $\Delta_L/J$. (b) Excitation spectra $E_{v,L}$ (black continuous lines) and $E_{v,R}$ (blue dot-dashed lines) as a function of anisotropy of $\Delta_L/J$ in the reverse bias. Vertical dashed lines, indicating dips in reverse current $\mathcal{J}_r$ of panel (a), occur where $E_{v,L} = E_{v,R}$ in panel (b). Other parameters: $N=6$, $J_m=0.1J$, $h=0$, $\Gamma=0.1J\hbar$. }%\dpc{Is $N=6$ or $12$?} \dpc{why are the lines red in the top panel? let us make them blue to be consistent also with Figs.2 and 3}}
\label{fig:fig4}
\end{figure}

Until this point we have considered segmented chains with integrable halves although the entire spin system is non-integrable. In this section, we investigate the robustness of the rectification to integrability breaking terms, by adding next-nearest neighbour tunneling and staggered fields to each half of the spin chain. To characterize the transport properties, we start by studying the magnetization profile, $\aver{\hat{\sigma}^z_n}$ as a function of the position $n$ in Fig.\ref{fig:sznnnstag}.
A linear ramp of the magnetization along the chain is seen for diffusive transport. When the chain is insulating, a step-like magnetization profile is obtained where each half of the chain has the magnetization imposed by the bath connected to them. Here we have considered a fairly short chain, but these insights can be applied to longer chains.

The effect of next-nearest neighbour tunneling terms is studied in Fig.\ref{fig:sznnnstag}(a, b) whereas the effect of staggered field is studied in Fig.\ref{fig:sznnnstag}(c, d). In forward bias, two things are important to notice: first, the magnetization at the edges does not reach the values $0$ and $-1$ which the bath is trying to impose. Hence there is always some current transfer with the baths; second, at the interface between the two half-chains, specifically between spins $n=4$ and $n=5$, the difference in the magnetization does not reach the full magnetization bias imposed by the baths. These two points suggest that an insulating behavior can be ruled out in the forward bias. The magnetization profile is very different for the reverse bias case as the anisotropy increases, as shown in  Fig.\ref{fig:sznnnstag}(b, d). For larger anisotropies we observe a step-like magnetization profile where both the magnetizations of the first and last spin become very close to the values the baths aim to impose, $-1$ and $0$, while the magnetization jump at the interface approaches $1$, indicating a very strong interface resistance.
These two panels are thus showing that the current in reverse bias is significantly suppressed at large anisotropies $\Delta_L$.
We have thus far shown that introducing integrability-breaking perturbations on each half of the chain does not prevent the emergence of a strong interface resistance in reverse bias, see \figref{fig:sznnnstag}(b) and (d), as the anisotropy in the left half increases. This is crucial as a flat magnetization profile in the right half of the system can be envisioned as originating from the ballistic nature of the right segment in \cite{balachandran_perfect_2018}. Our results clearly show that the nature of the transport does not play a role in the obtaining the step magnetization profile with a strongly resistive interface.

\begin{figure}[ht]
    \includegraphics[trim=0 0 0 0,clip, width=\linewidth]{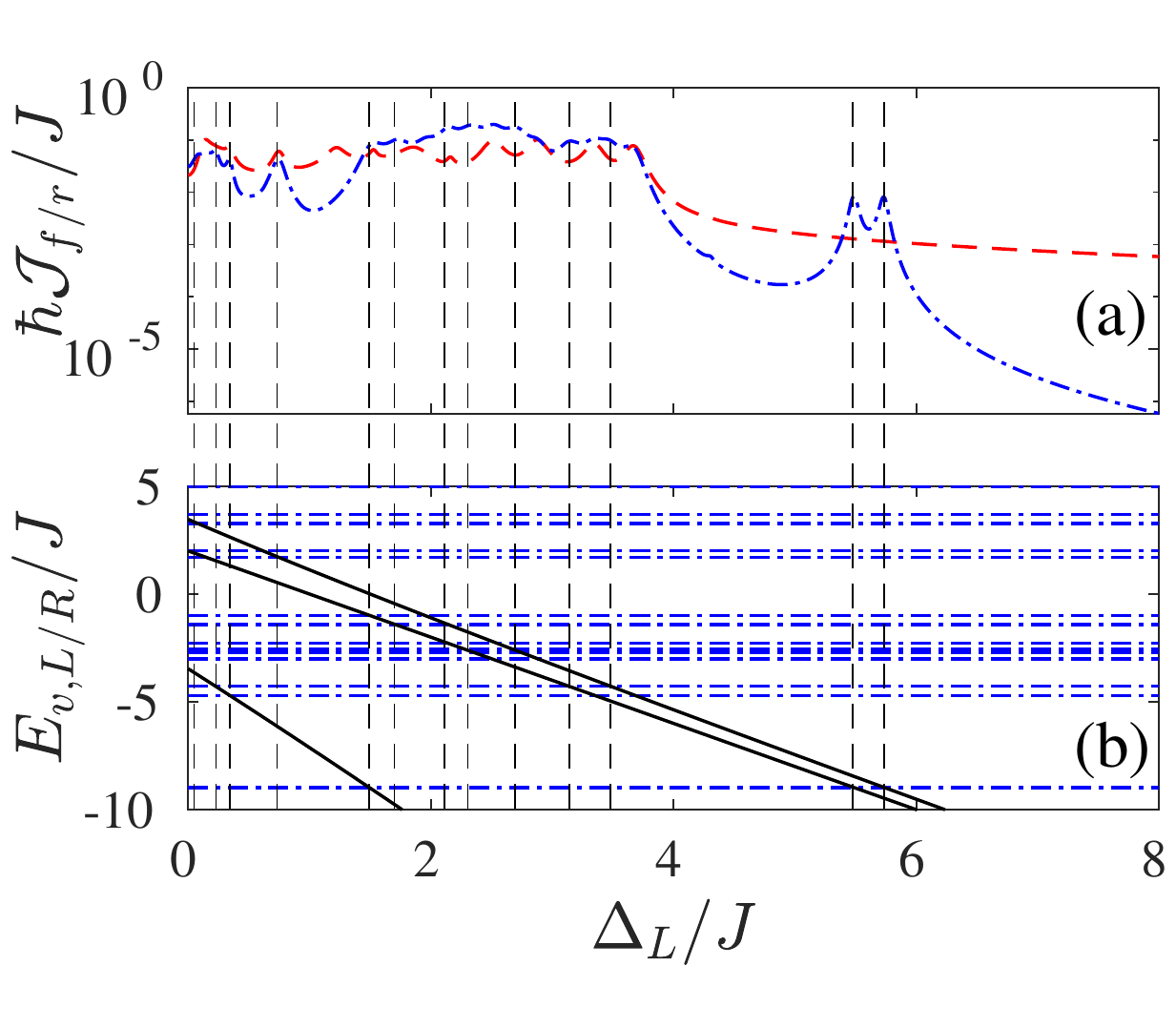}
\caption{Current $\mathcal{J}$ and excitation spectra $E_{v,L/R}$ against left anisotropy $\Delta_L$ with fixed staggered magnetic field perturbations $h=J$ and right anisotropy $\Delta_R/J=0.5$. (a) Forward current $\mathcal{J}_f$ (red dashed lines) and reverse current $\mathcal{J}_r$ (blue dot-dashed lines) against anisotropy on the left $\Delta_L/J$. (b) Excitation spectra $E_{v,L}$ (black continuous lines) and $E_{v,R}$ (blue dot-dashed lines) as a function of anisotropy of $\Delta_L/J$ in the reverse bias. Vertical dashed lines, indicating dips in reverse current $\mathcal{J}_r$ of panel (a), occur where $E_{v,L} = E_{v,R}$ in panel (b). Other parameters: $N=6$, $J_2=0$, $J_m=0.1J$, $\Gamma=0.1J/\hbar$. }%\dpc{Is $N=6$ or $12$?} \dpc{why are there red lines? let us make them blue to be consistent also with Figs.2 and 3}}
\label{fig:fig5}
\end{figure}

In \figref{fig:fig4} and \figref{fig:fig5}, we present a similar analysis as in Figs.\ref{fig:fig1_05enermatch} and \ref{fig:fig2_3enermatch}. More specifically, we study both the forward and reverse currents [in panels (a)], and the excitation spectra in each half [in panels (b)], as a function of $\Delta_L$.
Figs.\ref{fig:fig4} and \ref{fig:fig5}, the former with a next nearest-neighbour integrability breaking term and the latter with a staggered magnetic field, show that the underlying mechanism that causes insulating behavior in the reverse bias is still the opening of a gap in the excitation spectrum of the two halves of the chain as denoted by $E_{v,L}$ and $E_{v,R}$. At large anisotropies, the gap between $E_{v,L}$ and $E_{v,R}$ increases, resulting in a significant decrease of the reverse bias current. Moreover, all the peaks in the reverse bias current are due to crossings in the excitation spectra.
As a result, giant rectification can be obtained in non-integrable segmented halves provided anisotropy in one half is sufficiently large.

\begin{figure}[ht]
    \includegraphics[trim=0 0 0 0,clip, width=\linewidth]{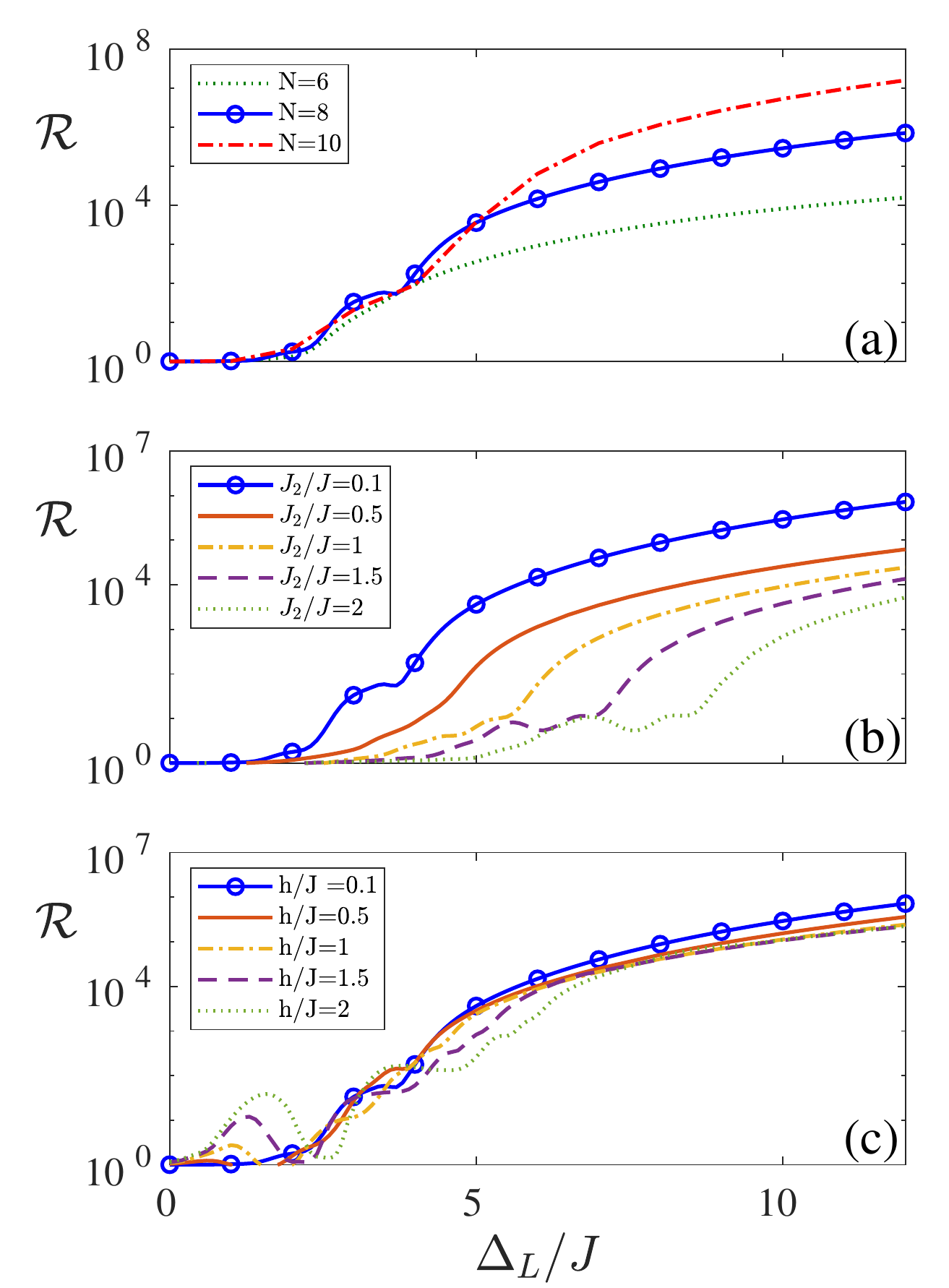}
\caption{Rectification $\mathcal{R}$ as a function of anisotropy $\Delta_L/J$ in a log-lin scale, for (a) systems of different chain lengths $N$ with next-nearest neighbour tunneling $J_2=0.1J$ and staggered magnetic field strength $h=0.1J$, (b) systems of chain length $N=8$ with different next-nearest neighbour tunneling at $h=0.1J$, (c) systems of chain length $N=8$ with different staggered magnetic field perturbations at $J_2=0.1J$. Blue lines with circle symbols indicate plots of the same system parameters with $N=8$, $J_2=0.1J$, $h=0.1J$ and $\Delta_R/J=0.1$. Other common parameters: $\Delta_R/J=0.1$, $J_m=J$, $\Gamma=J/\hbar$.}
\label{fig:Cnnnstag}
\end{figure}

For greater generality, we now consider both integrability breaking terms in the system, with the Hamiltonian given in Eq.(\ref{htotal}). In Fig.\ref{fig:Cnnnstag} we study the behavior of the rectification $\mathcal{R}$ for different systems sizes $N$, magnitude of integrability breaking terms, $J_2$ and $h$, and anisotropy in the left portion of the system $\Delta_L$. In particular, in Fig.\ref{fig:Cnnnstag}(a) we consider different system sizes while in Figs.\ref{fig:Cnnnstag}(b,c) we consider the effect of different magnitudes of integrability breaking terms for the spin chain Hamiltonian $\Ham$ in Eq.(\ref{htotal}); specifically we vary $J_2$ while keeping $h=0.1J$ in panel (b) and we vary $h$ keeping $J_2=0.1J$ in panel (c). In each panel we observe clearly that as the anisotropy $\Delta_L$ becomes larger, there is a significant increase in the rectification (note that Fig.\ref{fig:Cnnnstag} is in log-lin scale).

\begin{figure}[ht]
    \includegraphics[trim=0 0 0 0,clip, width=\linewidth]{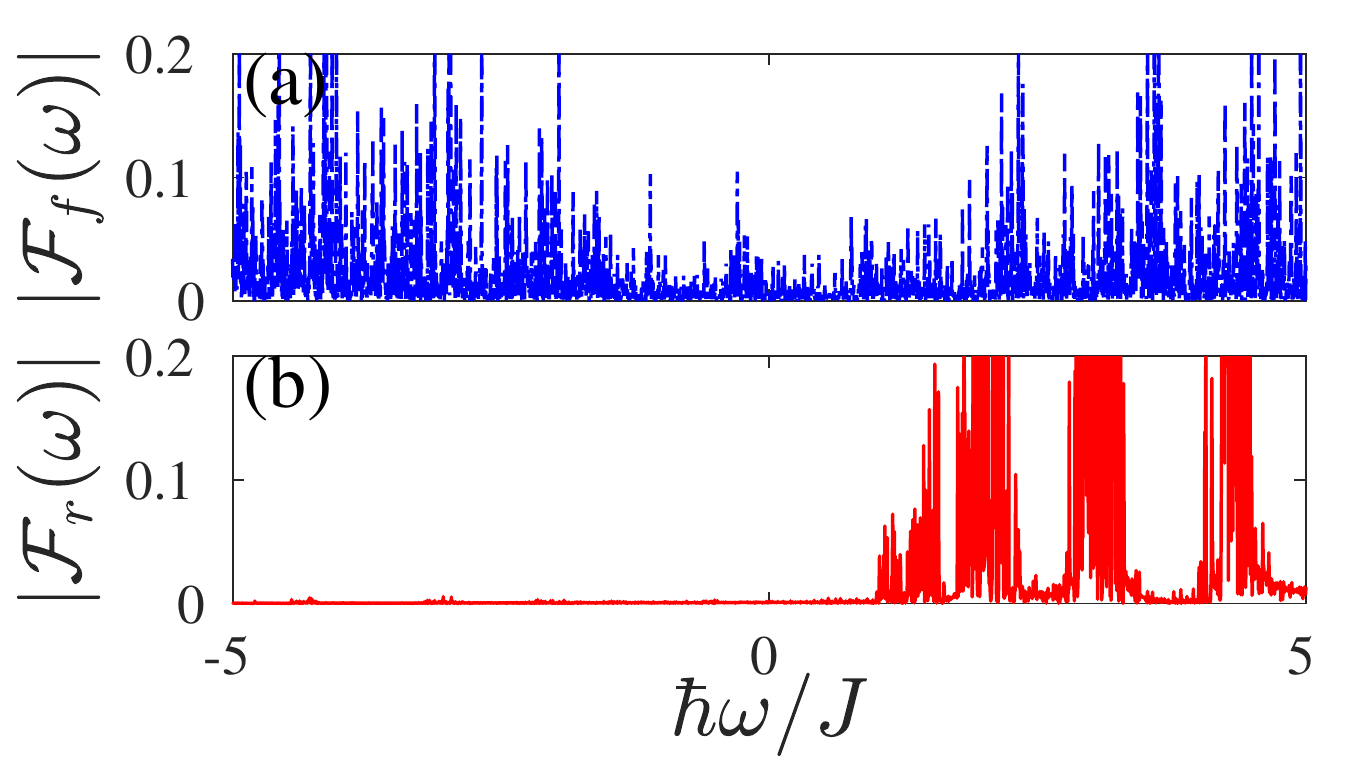}
\caption{Frequency response $|\mathcal{F}_{f/r}(\omega)|$ as a function of frequency $\hbar\omega/J$ for forward [blue dot-dashed lines in panel (a)] and reverse bias [red continuous lines in panel (b)] for a system size of $N=16$ with $J_2=0.1J$,  $h=0.1J$, $J_m=0.1J$, $\Delta_L/J=3.0$ and $\Delta_R/J=0.1$. The computations were done up to a time $t'=2000$ with $dt=0.05\hbar/J$.}

\label{fig:fig6}
\end{figure}

In \figref{fig:fig6}, we plot $\mathcal{F}(\omega)$ against $\omega$ for a system size of $N=16$, obtained from a Fourier transform of evolving Eq.(\ref{eq: 2timecorr}) for the two initial states $\hat{\rho}_{{\rm seg}, f}$ and $\hat{\rho}_{{\rm seg}, r}$. This is an exact parallel to \figref{fig:fig3_freqresp} however, this time, with integrability-breaking terms.
As expected, we see the emergence of an excitation gap in the reverse bias which is not present in the forward bias. This reaffirms that the mechanism for current rectification is due to the presence of an excitation gap and is independent of the integrable or non-integrable nature of each of the two half chains.

\section{Conclusion}\label{sec:conclusions}
Segmented XX-XXZ spin chains with large enough anisotropies can manifest giant rectification and even be perfect rectifier in the thermodynamic limit, as shown in \cite{balachandran_perfect_2018}. Here, by giant rectification we mean a rectification due to the fact that the system is an insulator in one bias, and present diffusive transport in the other. In this work we have shown that the emergence of giant rectification can be extended to generic integrable as well as non-integrable, magnetization conserving, interacting segmented spin chains. We have studied in greater detail the case in which both halves of the chain are interacting and integrable, and then we have also considered the effects of integrability breaking terms. In particular we have analyzed both an integrability breaking term which favors delocalization (a next-nearest-neighbor tunnelling term), and one which favors localization (a staggered magnetic field). In both case we have observed the emergence, in reverse bias, of an excitation gap which can turn the system into an insulator for finite system sizes. Interestingly, in forward bias transport is still possible, and hence the system shows giant rectification. In doing so we have clarified that the integrable character of each half of the segmented spin chain is not relevant, nor is whether the transport for one of the two halves is ballistic or diffusive. If one half of the chain has large enough anisotropy, an excitation gap in reverse bias will emerge, thus resulting in the insulating behavior for that bias alone.

There is however a fundamental difference with the model studied in \cite{balachandran_perfect_2018} for which, in the thermodynamic limit, there exist a finite critical anisotropy beyond which the system in reverse bias turns into an insulator. In the cases analyzed here, even without breaking of integrability, we have shown that the magnitude of anisotropy required for the system to become an insulator increases with the system size. This implies that while it is always possible to find a threshold anisotropy beyond which a finite-size system becomes insulating, there is no critical anisotropy in the thermodynamic limit.
While we have here studied the stability of this effect to integrability-breaking terms, future studies could investigate the role of other perturbations as, for instance, the role of an additional, dephasing, bath acting on each spin.

\begin{acknowledgments}
D.P. acknowledges support from the Ministry of Education of Singapore AcRF MOE Tier-II (Project No. MOE2018-T2-2-142). The computational work for this article was performed on resources of the National Supercomputing Centre, Singapore \cite{nscc}.
\end{acknowledgments}

\bibliography{bibliography}

\end{document}